# Phenomenological Simulation of Quantum Systems

John R Rankin

ProfessorJohn.Rankin@ANC.Vic.edu.au

**Abstract**

This paper describes an algorithmic system called SQT for the computer simulation of a wide class of quantum experiments on entangled particles. SQT maintains a hidden internal state for each particle and it provides an initialization process and measurement processes with visual outputs and a means of connecting these in software to make the simulator for a given experiment in Quantum Mechanics. The statistical behavior of the Quantum Mechanical system is replicated by incorporating the probabilities that are observed in the corresponding real-world quantum experiments. SQT is thus a tool to provide educational understanding of quantum systems and this tool removes the mystery surrounding entangled particles. It shows that only initial measurement probabilities and conditional probabilities of transition from the previous state and no earlier to the next state are sufficient inputs for the simulator to replicate all Quantum Mechanical experiments in the simulator and no other inputs such as complex numbers have to be entered into the simulator. The simulator, unlike the real-world quantum system, can display at any time the hidden internal sub-quantum state of each particle after which it can be clearly seen that there is no spooky interaction between entangled particles at all. SQT therefore provides a new interpretation of quantum phenomena that may well replace the Copenhagen interpretation founded by Niels Bohr and it answers Einstein's final question regarding whether Quantum Mechanics can be considered a complete theory or not.

**Key terms**

Quantum Mechanics, quantum systems, system initialization, measurement, entangled particles, hidden variables, HVT.

**Main Results and Assumptions**

The main result of this paper is the demonstration that the observed results of quantum experiments involving measurements on entangled particles can be fully simulated in their statistical behaviour in a computer software program that uses hidden variables. Other notable results are that no direct reference to a wavefunction and no use of complex numbers is required in the software. A further outcome is that the software only needs the number of particles and all ordinary probabilities and the conditional probabilities of all possible events in the experiments as inputs. The simulator and its algorithm as described in this paper applies to any number of particles even just for one particle but more typically for just two entangled particles. The main assumption is that quantum experiments can produce simultaneously not just two but even any number of entangled particles and further that each of the entangled particles can be separately measured in the standard variety of ways. Another assumption is that measurements on different particles will be done in a specific time order even if seeming to be done simultaneously in any real experiment. Thus by assumption, any experiment is a linear time sequence of measurements on various particles in the quantum system. The software is easily generated from the descriptions in this paper and additionally is available from the author.

## 1. Introduction

Computer Simulation of physical systems, most prevalent today in Games Technology [1], is a respected application of modern computers in the study of complex interacting systems with a long history of beneficial applications and results [2,3]. As examples, computer simulation is used in the training of airline pilots, the correct and efficient use of military equipment, the evaluation of war game strategies and in business optimization but today more profitably and more widespread, it is found in computer games and the entertainment industry. Whilst many different simulation languages exist in Computer Science, the concern in this paper is with the general conceptual application of computer simulation to the teaching and understanding of Quantum Mechanics. For the purposes of this work the class of quantum experiments of concern here are restricted to those static finite discrete systems where any number of entangled particles are created and then a sequence of measurements is done on the quantum system based on a given set of incompatible observables for the system. The aspects emphasized are the careful reinitialization of every quantum experiment and the careful scientific recording of what was done in each experiment and what the results of the experiment were. Subsequent statistical analysis of the recorded results is the final step in understanding the behavior of the quantum system.

Simulation design is often based on the finite state transition diagrams [1] and this is equally applicable to quantum systems where the real observable state of a quantum system (viz. stating what observable was last measured and what reading it had) rather than the complex quantum state $\psi$, is all we can observe and therefore limits all we can observationally know about a quantum system according to standard Quantum Mechanics [6,7,8]. The simulator described in this paper is based on SQT (standing for Sub-Quantum Theory) that uses observational states, probabilities and a hidden state array rather than the complex quantum state $\psi$ for the types of quantum experiments described in this instance. (A subsequent paper will show where $\psi$ occurs in SQT in other types of quantum experiments.) The number of possible observable states for a quantum system however grows rapidly with the number v of observables in the system and the number D of their possible readings per observable (where D is the maximum number of possible readings of any observable and in this work, D is assumed to be finite). Each reading is considered to cast the quantum system into a particular state because subsequent re-measurements for the same observable continue to return the same reading implying that the quantum system is projected into and staying in the same eigenstate of that observable with each measurement.

## 2. Simulation Components

To build a computer simulation of a quantum experiment it is necessary to define the software processes required for running the simulation of the experiment. Every quantum experiment requires the SI (System Initialization) process which initializes the quantum system for a repeated run of the experiment such that each experiment is a good and fair representation of what is done in the real world. The user must specify how many entangled quantum particles the SI will set up for each experiment. This is indicated as a parameter of the SI process so that SI(n) sets up a quantum system that creates n identical entangled particles. Next the user needs to specify the number of independent and incompatible observables that can be applied to the system per particle. These can be named for example as A, B, C etc. as needed for convenience in the simulator. Then the observables (equated to their corresponding measuring devices) will be applied to particular particles of the system. For example the experiment design specified as SI(3)+A(1)+B(2)+C(3)+A(1) means that first 3 entangled identical quantum particles will be created by SI, then observable A will be measured on the first particle then observable B will be measured on the second particle then observable C will be measured on the third particle and finally for this experiment observable A will be measured on the first particle again. The order of the measurements and the particles upon which the measurements are done is the significant feature of the experimental design that defines the experiment. With each measurement process the screen of the simulation software displays the value read by each measurement device and the results can be tabulated in a text file for the experiment so that the experimenter can review the experiment later and statistically analyse the results. In one approach to the design of the user-interface to the simulator, every observable (and its measurement device) could be an on-screen button and the user can either change the applicable particle number in an on-screen edit box for that observable or click on the observable's button to issue the measure command for that observable with the currently chosen particle number in the corresponding on-screen edit box. Figure 1 below shows the simulator screen interface for this approach in the case of two observables A and B with buttons to the left and particle number edit boxes to the right.

For the sake of uniformity, the readings for each observable are rescaled to being the natural numbers from 1 to D where D was entered by the user as the number of distinct readings per observable. If the number of possible readings for each observable is different from observable to observable, then take D as the maximum number of possible readings over all observables. If for example there are $v = 2$ observables A and B and observable A has 2 possible readings and observable B has 3 possible readings then take D as 3 and the 3$^{rd}$ possible reading for A is the same as the 2$^{nd}$ possible reading for A. All $v$ observables are by definition independent observables of the system in that the knowledge of the value of one observable gives no information on the current value for any other observable of the system. The observables may also be considered all incompatible observables such that the measurement of one observable disturbs the quantum system which changes the quantum state whereby the readings for previous measurements of other observables will no longer be valid. (The inclusion of compatible observables is discussed later.)

The design for a 2 observable simulator interface is shown in Figure 1. The user must enter all the settings for an experiment. This means firstly that the number of entangled particles n per experiment must be entered, then the number of possible readings, D, for each observable (which is the same for each observable) then the probabilities $p_{ij}$ and $P_{IJ}$ (which will be described in the next section) must be entered or the defaults accepted. After this setup an experiment can be done. For convenience, the setup can be saved to a file with the Save Settings button or loaded from a file with the Load Settings button. The Clear Settings button returns the simulator to the default settings. To do an experiment, first click the SI button. Suppose that next we want to measure observable A on entangled particle 1, then click the Measure A button. Suppose next that we need to measure observable B on entangled particle 2, then enter 2 in the edit box to the right of the Measure B button labelled "on particle" and after entering 2 click the Measure B button. Repeat this process for all the required measurements of the experiment in their desired order. The sequence of measurements and their results will be displayed in the memo box below the row starting "Experiment =". When all the measurements of the experiment have been done, the contents of the memo box can be copied to a text file or other document for saving and later review and statistical analysis. To repeat these experiments at any later time, the settings of the experiment can be saved to a file and loaded back in when we want to do more experiments on the same experimental setup. Repeated experiments will not in general give the same results as in any one batch of real quantum experiments but, by construction, the overall statistical performance has to be the same.

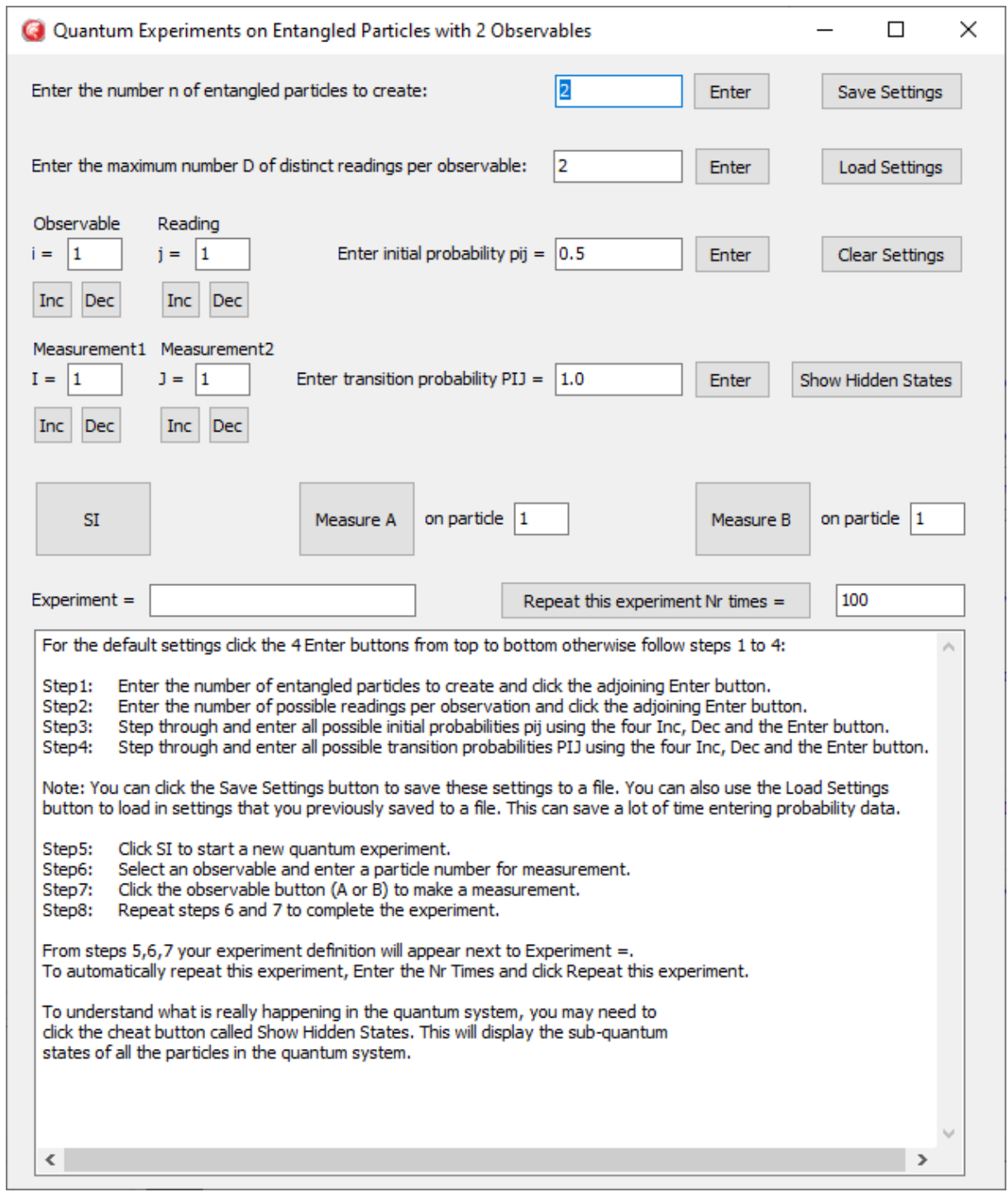

Figure 1: A user visual interface for simulating experiments on entangled particles where there are v = 2 observables.

The simulator interface design of Figure 1 is sufficient for experiments involving only 2 observables but a more general approach was subsequently needed. The preferred approach in this work is to allow the user to type in the experiment design instructions as a string, such as SI(3)+A(1)+B(2)+C(3)+A(1) above and then the process of running one or more experiments can be automated. Figure 2 below shows the user interface and the initial screen for this more general version of the simulator.

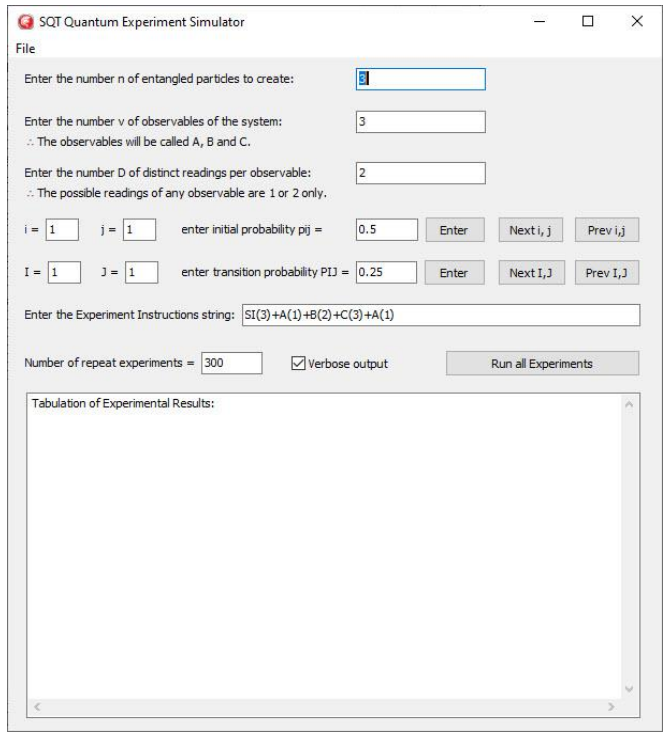

Figure 2: A user interface for the general SQT simulation experiments.

### 3. Entering the Probabilities

Before the simulation will run correctly, the user has to enter the first probabilities (also called the initial probabilities) $p_{ij}$ for each possible reading of each observable of the system and then enter all the second probabilities (also called the conditional or transition probabilities) $P_{IJ}$ from a particular reading of one observable to a particular reading of another observable. The initial probability $p_{ij}$ is the probability that if observable i, a number between 1 and v, is measured straight after SI is executed (i.e. directly after the quantum system was initialized) the result is the j'th possible reading for observable i where j is a number between 1 and D. Therefore, the software array $p_{ij}$ contains vD real numbers. The combination of the last measured observable number i and its corresponding reading number j defines the known last observable state of the quantum system denoted as I. So, I is a software variable consisting of two integers i = 1 to v and j = 1 to D and can be represented as an integer between 1 and vD. The quantum system can make a transition from the observable state I to the observable state J whenever there is a human interaction with the quantum system. The only human interactions after initialization of the quantum system is actual measurements made on the system. The transition probability $P_{IJ}$ is the probability that if the previous observable state of the system was I then after the interaction by another measurement the quantum system will be in observable state J. It follows that the software array $P_{IJ}$ contains $v^2D^2$ real numbers and in total the probability data needed for a quantum experiment consists of $vD + v^2D^2$ real numbers. This probability data comes from extensive real world experiments and is fairly tedious to gather and enter. It can alternatively come more easily from calculations from standard Quantum Mechanics. The $P_{IJ}$ data is a symmetric and square matrix with 1's down the diagonal. It is symmetric because the transition from state I to state J is the same as from state J to state I due to time reversibility in Physics. Since there are v observables and each has D possible readings we have vD initial probabilities to find and ½vD(vD-1) transition probabilities to find. If the smallest non-zero probability in the experiment is p then the number of experiments required to compute all the probabilities will be $N \geq \frac{vD(vD-1)}{2p}$. The first vD experiments will be of the form SI(1)+X(1) for the different observables X = A, B, C, …. For each of the v observables we need to do N experiments of the form SI(1)+X(1) to obtain $N_{ij}$, the number of occurrences of reading j on measuring the i'th observable X for j = 1 to D. Clearly $N = \sum_{j=1}^{D} N_{1j}$.

Repeating this for all v observables will fill up a table for v rows and D columns with frequency counts for each reading leading to an approximation for the initial probabilities of $p_{ij} = N_{ij}/N$. The remainder of the experiments will be of the form SI(1)+X(1)+Y(1) for different observables X and Y. This will fill up a large 2D table of values where there are vD rows and vD columns and each entry is a frequency count for each reading. This square array is similar to a Sudoku board for v = D = 3 and it consists of subsquares of size DxD with certain properties akin to 'magic squares'. The properties follow from the applicable principles of Physics which can be well-employed to reduce the tedium of the data entry for the setup of the simulator. To start with, the probability entries of each row must add up to 1 in every D x D subsquare and the probability entries of each column of a D x D subsquare must likewise add up to 1 due to the symmetry requirement. Secondly the D x D subsquares must be symmetric. Thirdly, the whole vD x vD table must be symmetric. The symmetry is due to the time reversal symmetry in the laws of Physics as mentioned above. Finally all entries along the main diagonal must be 1 and the other non-diagonal entries in the D x D subsquares down the main diagonal must be 0. This is based on the principle that immediate remeasurements should give the same answers. The case v = D = 2 is shown in Figure 3 below. The integers in parentheses in the table headings are the array indices I and J (observable states) each ranging from 1 to 4 = vD.

|   | A = 1 (J=1) | A = 2 (J=2) | B = 1 (J=3) | B = 2 (J=4) |
|---|---|---|---|---|
| A = 1 (I=1) | 1 | 0 | p | 1-p |
| A = 2 (I=2) | 0 | 1 | 1-p | p |
| B = 1 (I=3) | p | 1-p | 1 | 0 |
| B = 2 (I=4) | 1-p | P | 0 | 1 |

Figure 3. This is the state transition probability table $P_{IJ}$ for the case v = D = 2. Note that only one parameter p is required. (A = 1 means observable 1 called "A" was measured and the result was reading 1 and this is observable state I = (1,1) = 1, A = 2 means observable 1 called "A" was measured and the result was reading 2 and this is observable state I = (1,2) = 2, B = 1 means observable 2 called "B" was measured and the result was reading 1 and this is observable state I = (2,1) = 3, etc.)

It is clear from the figure above, that for v = D = 2 there is only one independent transitional probability component for $P_{IJ}$. It would be more efficient if for that case the simulator software would only ask for the one value p to be entered and then the simulator can fill out the full vD x vD $P_{IJ}$ array appropriately.

The case v = D = 3 is shown in Figure 4 below.

|  | A=1 (J=1) | A=2 (J=2) | A=3 (J=3) | B=1 (J=4) | B=2 (J=5) | B=3 (J=6) | C=1 (J=7) | C=2 (J=8) | C=3 (J=9) |
|---|---|---|---|---|---|---|---|---|---|
| A=1 (I=1) | 1 | 0 | 0 | p | q | 1-p-q | p' | q' | 1-p'-q' |
| A=2 (I=2) | 0 | 1 | 0 | q | 1-p-q | p | q' | 1-p'-q' | p' |
| A=3 (I=3) | 0 | 0 | 1 | 1-p-q | p | q | 1-p'-q' | p' | q' |
| B=1 (I=4) | p | q | 1-p-q | 1 | 0 | 0 | p" | q" | 1-p"-q" |
| B=2 (I=5) | q | 1-p-q | p | 0 | 1 | 0 | q" | 1-p"-q" | p" |
| B=3 (I=6) | 1-p-q | p | q | 0 | 0 | 1 | 1-p"-q" | p" | q" |
| C=1 (I=7) | p' | q' | 1-p'-q' | p" | q" | 1-p"-q" | 1 | 0 | 0 |
| C=2 (I=8) | q' | 1-p'-q' | p' | q" | 1-p"-q" | p" | 0 | 1 | 0 |
| C=3 (I=9) | 1-p'-q' | p' | q' | 1-p"-q" | p" | q" | 0 | 0 | 1 |

Figure 4. This is the state transition probability table $P_{IJ}$ for the case v = D = 3.

It is clear from the figure above, that for v = D = 3, 6 distinct transitional probability values need to be entered, namely p, q, p', q', p" and q". The simulator software could be made more efficient if for that case where v = D = 3, the simulator software would only ask for the 6 values p, q, p', q', p" and q" to be entered instead of the above 81 table entries. Since the most common case in Quantum Mechanics has v = D = 2 this paper will limit analysis to the v = D = 2 case and the v = D = 3 case will be taken up again in a later work.

4. **Design Variations For An Experiment**

The setup for an experiment is the number of entangled particles to create, n, the number of observables, v, the number of different readings, D, for any observable and all the initial and transitional probabilities of the system. For a given experimental setup there is an unlimited number of designs for experiments, each one conveniently represented by a string. For example the string "SI(1)+A(1)+B(1)" is the experiment where a single quantum particle is created by the system initializer SI and then the A observable is measured on the particle followed by the B observable measurement. Another design for an experiment on the same quantum system is SI(1)+A(1)+B(1)+A(1)+B(1). A more sophisticated experiment might have more particles and more observables such as

SI(3)+A(1)+B(2)+B(1)+C(3)+C(2)+A(1)+B(2).

Once the setup of an experiment is defined and once the initial and transitional probabilities for it have been entered then the different experiment design strings can be tested. The design strings are taken as instructions to the experimenter or to the simulator for running each experiment. The simulator design of Figure 2 makes it easy to run large numbers of repeated experiments on the same quantum system.

After the simulation is set up by the rules of sections 2 and 3 above, the user can specify how many times the experiment should be run. After setting this batch size, the user clicks the Run all Experiments button whereupon the simulation generates the results for all experiments and lists them in the on-screen Memo box for saving to a file. The user can choose via a CheckBox whether to create a verbose listing of the results or a tabulation of the numeric results only. By saving the results to a text file we can analyze them at a later time. The implementation of the code for this button involves reading the experiment design string from left to right. Firstly the 3 characters SI( are parsed and then the number in parentheses is scanned and the SI routine is invoked with the scanned number as parameter. Next the '+' character is skipped and the observable name e.g. 'A(' is scanned then the number in parentheses is scanned and the corresponding observable measurement routine is invoked with the scanned number as parameter. This scanning continues to the end of the experiment design string at which point the experiment is finished.

### 5. SI Routine Design and Observable Measurement Routine Designs

How is the SI routine implemented and how are the routines for measuring each observable implemented in the SQT simulator software? Here is where the essence of SQT contrasts with the standard interpretation of Quantum Mechanics. SQT works on a "sub-quantum state" distinct from the typical complex Hilbert space vector quantum state $|\psi>$ of standard Quantum Mechanics [6]. In SQT, SI prepares the system for the responses to all possible observable measurements in advance. This means giving each particle a v-tuple of potential next readings as hidden variables. This v-tuple is the sub-quantum state of the system. Every entangled quantum particle is assigned this same initial sub-quantum state v-tuple by the SI routine. The sub-quantum state cannot be directly seen by the experimenter (it is a "hidden variable" of the system) but by careful experimental design it can be deduced using the method that Einstein enunciated in his final 1935 paper [4] where the concept of entangled particles was first introduced. In the case v = 2, the sub-quantum state is an integer 2-tuple s = (a,b) where a is the reading that measuring A would give if it were measured straight after SI, and b is the reading that measuring B would give if instead of A it were measured straight after SI. Suppose that observable A is measured after SI then the A measurement process changes the sub-quantum state s to sub-quantum state s' = (a,b') and returns the reading as the natural number a between 1 and D. Clearly repeated measurements of observable A will always give the same reading a but in general the second natural number in the 2-tuple will change somewhat randomly every time the A reading is taken. After observable A is measured, the sub-quantum state has changed from (a,b) to (a,b') so a measurement of observable B will now return the reading b' and change the sub-quantum state again from s' to s'' = (a',b'). Every time there is a measurement by an independent observable in the system the sub-quantum state changes randomly in all components except the component corresponding to the observable measured. Although the component changes may seem random they are nevertheless subject to the transition probabilities, $P_{IJ}$, of the system.

It is essential to use both the first and second probabilities in implementing SI to create the initial SQT state of the system, s. The first probabilities are used to create the first component of s. The second probabilities are used to create the second component of s based on what the first component was. When v > 2, the conditional probabilities are also used to create all the subsequent components of s based on the value of the previous component. It is remarkable that the order chosen for the sequence of components is quite immaterial to the process. For example if v = 3 then for implementing SI we want to determine a, b and c in s = (a,b,c) so we could generate b by the first probabilities $p_{Bb}$ and then generate a by the second probability $P_{BbAa}$ and then generate c by the second probability $P_{AaCc}$. Alternatively, we could generate c by the first probabilities $p_{Cc}$ and then generate b by the second probability $P_{CcBb}$ and then generate a by the second probability $P_{BbAa}$. In the actual implementation of the SI routine, SI will generate reading a by the first probabilities $p_{Aa}$ and then generate reading b by the second probability $P_{AaBb}$ and then generate reading c by the second probability $P_{BbCc}$ to generate the hidden sub-quantum state 3-tuple (a,b,c) which is stored the same for each entangled particle created. Any chosen sequence of component generation is as good as any other and produces the same statistical results in experiments as software tests of this have shown. Next we will look at how to generate a component from the first probabilities and after that how to generate a component from the second probabilities.

Let $p_{ij}$ be the probability of the j'th possible reading (of value j between 1 and D) for the i'th observable immediately after system initialization where i goes from 1 to v. Note that

$$\sum_{j=1}^{D} p_{ij} = 1$$

A simple way to use $p_{ij}$ to implement the allocation of an initial value for component 1 say, is to imagine running a large number N of experiments of the form SI(1)+A(1) where reading j appears with (approximate) frequency $N_j = Np_{1j}$ (the nearest integer part of the product being taken here). Now using a uniform pseudo-random number generator, generate a random number r between 1 and N. The A component of the initial state s is that integer j such that

$$\sum_{k=1}^{k=j-1} N_k < rN \leq \sum_{k=1}^{k=j} N_k$$

(In the case where j = 1 the left hand sum is replaced with 0. In the case where j = D the right hand sum is replaced with N.) By rN is meant the nearest integer to the floating point product. Equivalently dividing the above inequations through by N, we need to form the sequence of the cumulative sums $S_{ij}$ of the initial probabilities $p_{ik}$ for fixed i and for k = 1 to j:

$$S_{ij} = \sum_{k=1}^{j} p_{ik}$$

Then the component value $i_2$ for observable i is the minimum j such that a uniform PRN (pseudo-random number) is less than $S_{ij}$.

A similar implementation is needed for each observable measurement based on the possible sub-quantum state transition probabilities. Here we use the second probabilities $P_{IJ}$ for I, J = 1 to vD and using J = ($j_1$,$j_2$) with $j_1$ from 1 to v and $j_2$ from 1 to D

$$\sum_{j_2=1}^{D} P_{IJ} = 1$$

$$i_1 = (I-1)\text{div}D + 1 \qquad j_1 = (J-1)\text{div}D + 1$$
$$i_2 = (I-1)\text{mod}D + 1 \qquad j_2 = (J-1)\text{mod}D + 1$$
$$I = (i_1-1)D + i_2 \qquad J = (j_1-1)D + j_2$$

(Here div is the integer division function which truncates real division by removing the fractional part.) The corresponding observable numbers $i_1$ and $j_1$ are the incremented quotients of the decremented indices, I-1 and J-1, on division by D as given by the integer div function (that is $i_1 = (I-1)\text{div}D + 1$, $j_1 = (J-1)\text{div}D + 1$) and the corresponding readings are $i_2$ and $j_2$, which are the remainders as given by the integer mod function modulo D incremented or more simply by $i_2 = I - 2i_1 + 2$. For example in the simple case of v = D = 2, we have $P_{11}$ = 1, $P_{12}$ = 0, $P_{13}$ = p, $P_{14}$ = 1-p, $P_{21}$ = 0, $P_{22}$ = 1, $P_{23}$ = 1-p, $P_{24}$ = p, $P_{31}$ = p, $P_{32}$ = 1-p, $P_{33}$ = 1, $P_{34}$ = 0, $P_{41}$ = 1-p, $P_{42}$ = p, $P_{43}$ = 0 and $P_{44}$ = 1 as shown in Figure 3 above. For instance, the I = 2 on $P_{23}$ refers to observable $i_1$ = 1 (i.e. A) corresponding to $i_1$ = (2-1) div D +1 = 1, with reading $i_2$ = 2 corresponding to $i_2$ = I-$2i_1$+2 = 2 = (2-1) mod D+1, and the J = 3 refers to observable $j_1$ = 2 (i.e. B) corresponding to $j_1$ = (3-1) div D + 1, with reading $j_2$ = 1 corresponding to $j_2$ = J – $2j_1$ + 2 = (3-1 mod D) + 1 = 1. (Thus $P_{23}$ can equivalently be written more clearly as $P_{A2B1}$ etc.)

To choose a transition we would form the cumulative sums:

$$S_{IJ} = \sum_{K=1}^{J} P_{IK}$$

As before, the chosen transition J from the previous observational state I is the minimum J such that a uniform PRN is less than $S_{IJ}$. From J we derive $j_1$ and $j_2$ as above and these integers tell us the next component number $j_1$, and the value, $j_2$, to put at that component in the SQT system state s.

6. **General Case for v = D = 2**

For the v = D = 2 case we saw that two initial probabilities $p_1$ and $p_2$ are required and one transition probability $p_3$ is required (see Figure 3). Here $p_1$ is the probability that if A is measured on particle 1 after SI the answer is A(1) = 1, $p_2$ is the probability that if B is measured on particle 1 after SI the answer is B(1) = 1 and $p_3$ is the probability that if A is measured on particle 1 with reading A(1) = 1 (observational state I = 1 for particle 1) and then next B is measured on particle 1 it will have reading B(1) = 1 (observational state J = 3 for particle 1). To implement SI(n) we need to run the uniform pseudo-random number generator to obtain a random real number, PRN, between 0 and 1. The SI procedure will create a sub-quantum state which is a 2-tuple s = (a,b) that has to conform to the given initial probabilities defined by $p_1$ and $p_2$ and the transition probability $p_3$. The rules are:

Obtain a new PRN. If PRN ≤ $p_1$ then a = 1 else a = 2.

Obtain a new PRN. If a = 1 then (if PRN ≤ $p_3$ then b = 1 else b = 2)

　　　　　else (if a = 2 and PRN ≤ 1-$p_3$ then b = 1 else b = 2).

(The second rule requires nested if-then-else statements.) This makes the new sub-quantum state s = (a,b) which must be stored for each of the n entangled particles (as n hidden variables of the system). SI dynamically allocates an n-element array of integer type to store the state s for each of the n entangled particles. Throughout the SQT simulator the observational state J is for implementation convenience converted to an integer and back via the mapping 1 = (1,1), 2 = (1,2), 3 = (2,1) and 4 = (4,4).

To implement the measuring device for an observable such as A we follow a similar procedure. Firstly, we need to look at the parameter e on the procedure A(e). This tells the software code for measuring A, the sub-quantum state s of which entangle particle to look at, i.e. which one of the n copies of the n sub-quantum states to look at. Suppose we find that the stored sub-quantum state for particle e is s[e] = (a,b). Then A as the first observable will return the reading a to the experimenter (the user) and the A measurement process will alter the second parameter of the sub-quantum state, the b value, giving the new state s'[e] = (a,b') by the rule:

Obtain a new PRN. If PRN ≤ $P_{1x21}$ then b' = 1 else b' = 2.

This makes the new sub-quantum state s' for particle e which must be stored with particle e and now particle e is no longer entangled with the other particles. No flag is required to be stored in the simulator to say whether a particle is entangled or not. Any two particles say p and p' are 'entangled' if they both still have the same initial sub-quantum state v-tuple unchanged, i.e. s[p] = s[p']. However the sub-quantum state is a 'hidden variable' in the sense of Einstein and it is not directly accessible to the experimenter. Therefore the experimenter can only know which particles are still entangled by remembering which have not yet been measured by any device. (The measurement process destroys entanglement.) Note also that the simulation system does not need to keep a record of which measurement was last done so that the appropriate transition probability would be applied. This is because SQT applies the appropriate transitive probabilities pre-emptively.

To implement the measuring device for the second observable B we follow a similar procedure. Firstly we need to look at the parameter e on the procedure B(e). This tells the code for B which of the n copies of the n sub-quantum states to look at. Suppose we have the appropriate state of that particle as s = s[e] = (a,b). Then B as the second independent observable will return the reading b to the experimenter and the B measurement process will alter the first parameter of the sub-quantum state, the a value, giving the new state s' = (a',b) by the rule:

Obtain a new PRN. If PRN ≤ $P_{112y}$ then a' = 1 else a' = 2.

This makes the new sub-quantum state s' for particle e which must be stored with particle e and now particle e is no longer entangled with the other particles. Again, no flag is required to be stored to say whether a particle is entangled or not. Any two particles are 'entangled' if they still have the same sub-quantum state tuple. Having the same sub-quantum state tuple by accident does not count as entangled particles either (which can often happen for such low values of v and D). However, the sub-quantum state is a 'hidden variable' in the sense of Einstein and it is not directly accessible to the experimenter. Note also that the simulation system does not need to keep a record of which measurement was last done in order to know what the appropriate transition probability would be to apply. In all these processes, the software does not reveal any subquantum state (a,b) to the user and the only information that the user gets is the measurement reading value of a or b as the case may be after which the other component is scrambled.

By ensuring that SI generates the first component of the SQT system state by the first probability table and the second component by the second probability table we are assured that the statistics of the generated simulation experiment results will be in accord with the results of Quantum Mechanics and with real physical experiments. The following screen shots, Figures 5a and 5b below, show a test of this.

```
Quantum Experiments on Entangled Particles with 2 Observables               — □ ×

Enter the number of entangled particles to create:        [2]    [Enter]    [Save Settings]

Enter the maximum number of distinct readings per observable: [2]  [Enter]  [Load Settings]

Observable   Reading
i = [1]      j = [1]       Enter initial probability pij = [0.5]   [Enter]   [Clear Settings]
[Inc] [Dec]  [Inc] [Dec]

Measurement1  Measurement2
I = [1]       J = [1]      Enter transition probability PIJ = [1]  [Enter]   [Hidden State]
[Inc] [Dec]   [Inc] [Dec]

        [ SI ]              [Measure A]  on particle [1]         [Measure B]  on particle [1]

[SI(2)+A(1)+B(2) test. Repetitions =]   [100000]

Starting a new quantum experiment.
The number of observables = 2
The maximum number of possible readings per observable = 2
The number of particles = 2
Settings file loaded = Aspects Experiment.SQT

SI(2)+A(1)+B(2) with 100000 repetitions gives:

N11 = 43114      p11 = 0.43
N12 = 6911       p12 = 0.07
N21 = 7039       p21 = 0.07
N22 = 42936      p22 = 0.43
```

Figure 5a. Test run of experiment SI(2)+A(1)+B(2) with 100,000 repetitions using the probabilities from Aspect's experiment.

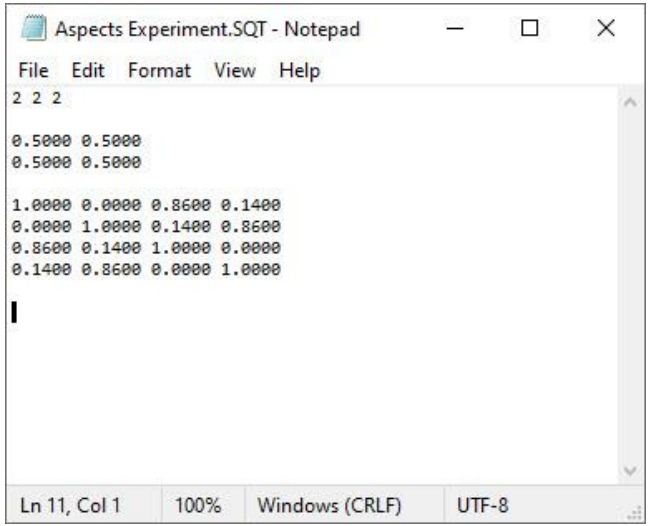

Figure 5b. The values of v, D, n, the first probability table and the second probability table for Aspect's experiment.

## 7. Results

A run of the simulator for v = D = 2 with verbose result reporting is shown in Figure 6 below.

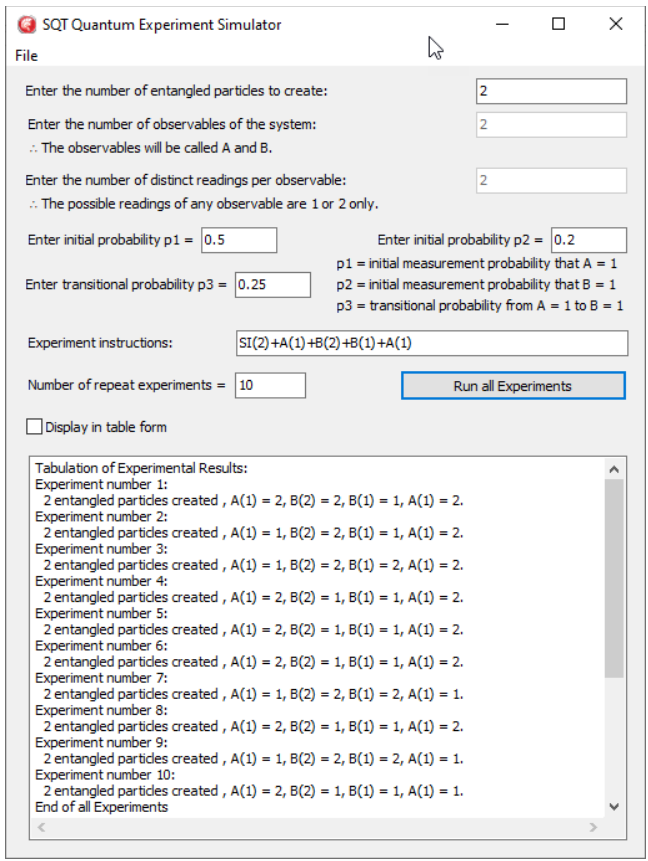

Figure 6. A sample run for the case n = v = D = 2.

In this experiment we see that only 10 experiments were done. Normally we would do a lot more but a small number is useful in checking what the software is doing. Even though the 10 experiments were identical, the results recorded and displayed in the above figure are different because quantum experiments are statistical in nature. In each experiment 2 entangled quantum particles were created. Observable A was measured on the first particle then observable B was measured on the second particle and then on the first particle and finally A was measured on the first particle again. The first A measurement has variable results which in the long run are 50/50 between the 1 and 2 because $p_1$ = 0.5. We notice that the second A measurement of the first particle is often not the same as the first A measurement result. This is because between the two measurements of A, observable B is measured on the same particle. Since A and B are incompatible observables, the B measuring device randomly messes up the next potential reading for observable A. The B measurement on the second particle also varies since with the given initial probabilities the system initializer process SI will setup different initial B states. The B(1) and B(2) measurements are different because the A(1) measurement changed what B would next be measured as on particle 1 whilst not changing what B would be measured as on particle 2.

The simulator also prints out some statistics at the end of the table of results. The statistics for the above run are shown below in Figure 7.

```
Statistics Report:
There were 4 measurements per experiment.
Measurement 1 gave 1 4 times.
Measurement 1 gave 2 6 times.
Measurement 2 gave 1 5 times.
Measurement 2 gave 2 5 times.
Measurement 3 gave 1 7 times.
Measurement 3 gave 2 3 times.
Measurement 4 gave 1 3 times.
Measurement 4 gave 2 7 times.
End of Statistics Report.
```

Figure 7. Statistics reported by the simulation for the run shown in Figure 5a.

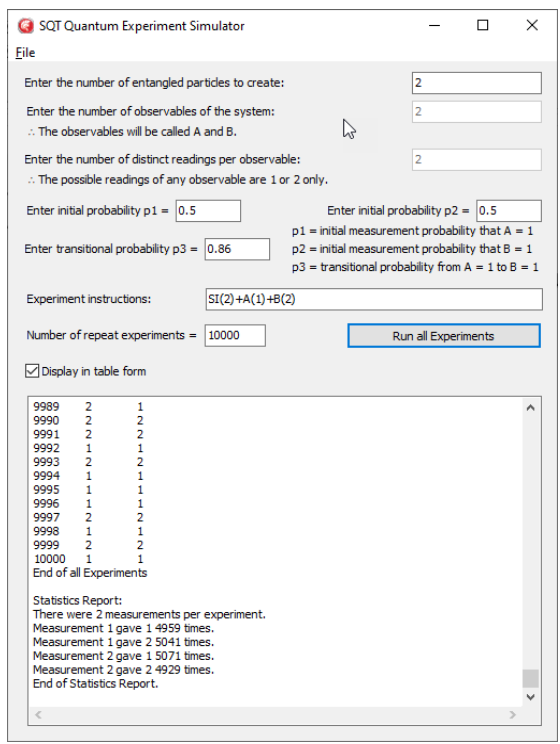

Figure 8. Simulation results for the Alain Aspect quantum experiment.

Another type of statistical analysis is to check whether the distribution of results is in accordance with real quantum experiments or not. To do this a much larger number of experiments is required. For the case of Alain Aspect's Bell test experiments we have n = 2, v = 3, D = 2 initial probabilities $p_1 = p_2 = 0.5$ and transitional probability $p_3 = 0.86$. The test run is shown in Figure 8 above for 10,000 replications of the experiment SI+A(1)+B(2).

The tabulated raw results were saved in a text file of 10,000 lines with 3 integers per line, the experiment number, the A(1) reading and the B(2) reading, and then run through another program to determine the distribution by counting the number of times the results were 1, 1 as $N_{11}$ or 1, 2 as $N_{12}$ etc. as seen in Figure 9 below.

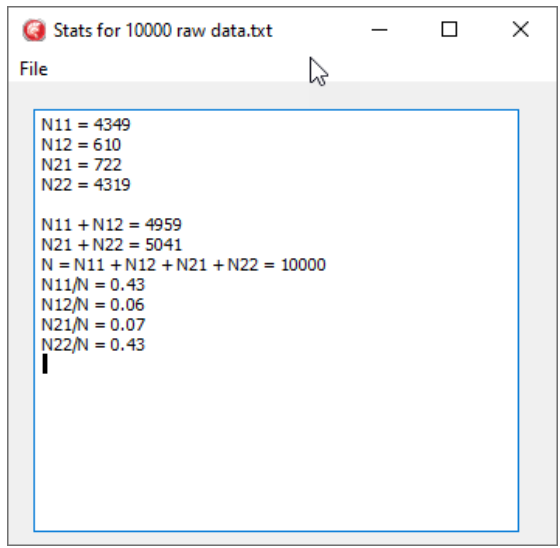

Figure 9. Analysis of the run of the Alain Aspect quantum experiment.

This analysis shows the expected distribution in the experimental results in accordance with the real quantum experiments.

8. Discussion

The simulation system described in this paper is designed to exactly match the experimentally observed behavior of a class of quantum systems and is powered by an algorithm called SQT. However the model takes us further by showing a new way of interpreting the behavior of quantum systems which is called Sub-Quantum Theory (SQT for short). It provides a parallel interpretation of Quantum Mechanics to the original Copenhagen interpretation given by Niels Bohr [5] in his arguments with Albert Einstein [4]. In the Copenhagen interpretation the D distinct possible readings for each of the v observables suggested the need for a D-dimensional Hilbert space with the state of the quantum system as a complex vector $|\psi\rangle$ in this admittedly non-physical and unobservable though useful conceptual space. For n entangle particles, the composite state vector is a Cartesian product of n such vectors from n copies of the D-dimensional Hilbert space. In Quantum Mechanics, the D different readings for any system observable applied to one particle, result from its measuring device projecting the particle's state vector into a postulated orthonormal basis for its complex Hilbert space. Each observable therefore corresponds to a different orthonormal basis obtained by a D-dimensional rotation from any other. When an observable is measured on a particle, the measuring device for that observable effectively projects the state of the system as a Hilbert space vector onto one of the basis vectors for that observable corresponding to the reading obtained (which is the eigenvalue). The norm (length squared) of the projected complex state vector relative to the norm of the unprojected vector is the probability of the transition. In the Copenhagen interpretation the initial state $|\psi>$ is a superposition of all possible states of an orthonormal basis and this represents a non-physical state i.e. a state with no determinate value of the observable and only when a measurement is done on the system does the system enter a knowable physical state. This is often stated as a quantum particle not 'knowing' what value it will have before measurement (or even extremely, that the particle doesn't 'exist' until it is measured).

On the other hand, in SQT we have to do a large number N of experiments to find the probabilities of each observable initial reading and the probabilities of the transitions between the different possible observational states of the quantum system. These probabilities could also come from standard Quantum Mechanics which likewise ultimately derives the probabilities from experiment. SQT models the quantum system phenomenologically without resorting to complex Hilbert space vectors. It posits a hidden state for each particle called the particle's sub-quantum state which is a v-tuple where the i'th component is a natural number between 1 and D corresponding to the next reading of observable i should the i'th observable be read next for that particle. The i'th measurement device scrambles all components of the particle's sub-quantum state but the i'th component and returns the i'th component to the experimenter as the reading for that measurement. In SQT therefore each particle does 'know' in advance what value it will have for any observable measured on the particle (that the experimenter may freely choose). SQT is also a 'local theory' in that the measurement of any one particle has no effect on what some other particle might be subsequently measured as whether nearby or far-off: they each have an independent sub-quantum state v-tuple. In SQT there is no mysterious influence flowing between so-called entangled particles. It is no longer even necessary to call the particles 'entangled' except in recognition that their physical properties are initially (before any measurements) tied together by the relevant conservation laws. The conservation laws get modified when the measurement instruments interact with the system. Yet by construction, the SQT simulation system does replicate the statistical behavior of real quantum systems of entangled particles. By 'realism' in Physics, Einstein [4] meant that existence is not dependent on observation or measurement. SQT has this kind of 'realism' since the reading that any observable will give on a particle is definite and 'exists' (inside the hidden SQT state v-tuple) before the measurement is ever done (though being hidden from the observer). In SQT the Moon exists even when you don't look at it.

The SQT hidden state (a,b) for a quantum system with v = 2 incompatible observables A and B is reminiscent of the classical state of a particle in one-dimension as a point in phase space (x,p) with observables X and P measuring position and momentum in the thought experiment of Einstein [4]. (Although x and p are continuous variables and SQT was described in this paper for discrete variables a and b, SQT is clearly extendible to continuous variables though this could require significant changes for use in simulator software. The distinction that observables A and B have a finite and discrete range of possible readings whilst X and P have a continuous and infinite range serves only to make the implementation of the software simulator easier.) Einstein argued [4] that for a quantum system when X is measured x is returned and p is disturbed by Δp to produce a new phase state point (x,p') where p' = p + Δp. Alternatively, when P is measured p is returned and x is disturbed by Δx to result in a new phase state point (x',p) where x' = x + Δx and ΔxΔp ≥ ½ℏ. If we include compatible observables such as for a particle moving in 2D, we would have the phase state as a 4-tuple s = $(x,y,p_x,p_y)$ with observables X, Y, $P_x$ and $P_y$ and the only non-commuting pairs are $[X,P_x]$ and $[Y,P_y]$. Then measuring X returns x and disturbs the state only to $(x,y,p_x',p_y)$ or measuring $P_x$ returns $p_x$ and disturbs the state only to $(x',y,p_x,p_y)$, or measuring Y returns y and disturbs the state only to $(x,y,p_x,p_y')$ or measuring $P_y$ returns

$p_y$ and disturbs the state only to (x,y',$p_x$,$p_y$). These similarly add unenlightening complexities to the software implementation and so were avoided in this work.

9. Conclusions

With modern software tools it is not difficult to create simulation software to replicate the behavior of classes of experiments on entangled particles in Quantum Physics. For simulator verification and validation it is necessary to check that the statistical distribution of observable readings over a large number of experiments replicates what is seen in nature. This has been done in the SQT simulation software which has been shown to copy nature. The theoretical modelling of these experiments has led to a new interpretation of Quantum Mechanics beyond the original Copenhagen interpretation promoted by Niels Bohr. This new interpretation, called SQT, is both local and has realism as defined by Einstein and therefore seems to be the view of Quantum Mechanics that Einstein sought for in his long discussions with Niels Bohr. In other words, it appears that SQT completes Quantum Mechanics in the way that Einstein sought for a complete and realistic theory. How SQT which is effectively an HVT, performs in a Bell experiment will be presented in a subsequent paper.